\documentclass{elsart}
\usepackage{epsfig}
\newenvironment{pacs}{\vspace{0.5cm}\begin{flushleft}{\em P.A.C.S.:\/}}%
{\end{flushleft}}
\begin{document}
\begin{frontmatter}
\title{Effective field approach to the Ising film in a transverse field}
\author[Trieste,Napoli]{Luca Peliti}
\author[Trieste,Napoli,Meknes]{Mohammed Saber}
\address[Trieste]{Abdus Salam International Centre for Theoretical Physics,\\
P.O.B. 586, I-34100 Trieste, Italy}
\address[Napoli]{Dipartimento di Scienze Fisiche and Unit\`a INFM\\
Universit\`{a} ``Federico II'', Mostra d'Oltremare, Pad.~19, I-80125 Napoli, Italy}
\address[Meknes]{D\'epartement de Physique, Facult\'e des Sciences, Universit\'e
``Moulay Isma\"\i l'',\\
B.P. 4010, Mekn\`es, Morocco\thanksref{Mail}}
\thanks[Mail]{Address for correspondence.}

\begin{abstract}
We study the phase transitions of the spin-$\scriptstyle{\frac{1}{2}}$ 
Ising film in a transverse field within the framework of the effective field theory.
We evaluate the critical temperature of the film as a function of the exchange
interactions, the transverse field and the film thickness. We find
that, if the ratio of the surface exchange interactions to the
bulk ones $R=J_{\rm s}/J$ is smaller than a critical value $R_{\rm c}$, 
the critical temperature $T_{\rm c}/J$ of the film is smaller than the 
bulk critical temperature $T_{\rm c}^{\rm B}/J$ and approaches $T_{\rm c}^{\rm B}/J$
as $R$ increases further.
On the other hand, if $R>R_{\rm c}$, $T_{\rm c}/J$ is larger than both
the bulk $T_{\rm c}^{\rm B}/J$ and the surface $T_{\rm c}^{\rm S}/J$ 
critical temperatures of the corresponding semi-infinite system, and
approaches $T_{\rm c}^{\rm S}/J$ as $R$ increases further.


\end{abstract}

\end{frontmatter}

\section{Introduction}
During the last years much effort has been directed towards the study of
critical phenomena in various magnetic layered structures, ultrathin films
and superlattices \cite{1,2,3,4}. The basic theoretical problem is the examination
of the magnetic excitation and the phase transitions in these systems.
Of these, magnetic films are very important both from the theoretical
and the experimental standpoints \cite{5,6}, and can be studied as
models of the magnetic size effect [7] and as quasi-two-dimensional systems.
The magnetic and phase transition properties of semi-infinite Ising
systems have been investigated for many years.

The surface magnetism of these systems is very interesting \cite{8,9,10,11,12,%
13,14,15,16,17}. 
It exhibits different types of phase transitions associated with the 
surface; if the ratio $R$, ($R=J_{\rm s}/J$), is greater than a critical 
value $R_{\rm c}=(J_{\rm s}/J)_{\rm crit}$, the system
may order on the surface before it orders in the bulk. 
As the temperature is lowered, the system undergoes
two successive transitions, namely the {\em surface\/} and {\em bulk\/} 
phase transitions, whose critical temperatures are
called the surface ($T_{\rm c}^{\rm S}$)  and the
bulk ($T_{\rm c}^{\rm B}$) critical temperatures respectively.
On the other hand, if the ratio $R$ is less than $R_{\rm c}$, the whole
system becomes ordered at the bulk transition temperature 
$T_{\rm c}^{\rm B}$.

Magnetic excitations in superlattices were considered in numerous papers
(see, e.g., ref.~\cite{18} for a brief review). However, less attention
has been paid to critical behavior, and in particular to critical
temperatures in superlattices. Ma and Tsai \cite{19} have studied the
variation with the modulation wavelength of the Curie temperature for a
Heisenberg magnetic superlattice. Their results agree qualitatively with
experiments on Cu/Ni films \cite{20}. Superlattice structures composed
of alternating ferromagnetic and antiferromagnetic layers have been
investigated by Hinchey and Mills \cite{21,22}, using a localized spin
model. A sequence of spin-reorientation transitions are found to be
different for superlattices with the antiferromagnetic component
consisting of an even or odd number of spin layers.

Fishman et al.~\cite{23} have discussed,
within the framework of the Ginzburg-Landau formalism, the static and 
dynamical properties of a periodic multilayer system formed of two different 
ferromagnetic materials. They have 
computed the transition temperature and the spin-wave spectrum. On the other 
hand, the Landau formalism of Camley and Tilley \cite{24} has been applied to 
calculate the critical temperature in the same system \cite{25}. Compared to 
ref.~\cite{23}, the formalism of ref.~\cite{24} appears to be more general because 
it allows for a wider range of boundary conditions and includes the sign 
of exchange coupling across the interface.

For more complicated superlattices with arbitrary number of 
different layers in an elementary unit, Bern\'as \cite{26} has derived some 
general dispersion equations for bulk and surface magnetic polaritons. 
These equations have been then applied to magnetostatic modes and to retarded wave 
propagation in the Voigt geometry \cite{27}.

On the other hand, with the developement of modern vacuum 
science and in particular the epitaxial growth technique, it is possible to 
study experimentally the magnetic properties of low dimensional systems. 
For example, by depositing magnetic atoms on the top of non magnetic 
substrates, the thickness dependence of the critical temperature of 
ultrathin films of Gd on W(110) \cite{28} and of Fe on Au(100) \cite{29}, 
has been measured.

Our aim in this paper is to study the phase diagrams of an Ising 
(spin-$\scriptstyle{\frac{1}{2}}$) film in a transverse field within the framework 
of the effective field theory \cite{30}. This technique is believed to give 
more exact results than those of the standard mean-field approximation.
In section 2 we outline the formalism and derive the equations
that determine the layer magnetizations, the average magnetizations 
and the critical temperature of the film as functions of temperature, 
exchange interactions, transverse fields and film thickness.
The phase diagrams of the film are discussed in section 3.
The last section is devoted to a brief conclusion.

\section{Formalism}
We consider a spin-$\scriptstyle{\frac{1}{2}}$ Ising film of $L$ layers on a simple cubic
lattice with free surfaces parallel to the $(001)$ plane, submitted to a 
transverse field. The Hamiltonian of the system is given by
\begin{equation}
H=-\sum_{{<}ij{>}}J_{ij}\sigma_{i}^{z}
\sigma_{j}^{z}-\sum_{i}\Omega_{i}\sigma_{i}^{x},\label{eq1}
\end{equation}
where $\sigma_{i}^{z}$ and $\sigma_{i}^{x}$ respectively denote the $z$ and $x$ components
of a quantum spin $\vec{\sigma}_{i}$ of magnitude $\sigma=\frac{1}{2}$ at 
site $i$, $J_{ij}$ is the strength of the exchange interaction between
the spins at nearest-neighbor sites $i$ and $j$, and $\Omega_{i}$ represents 
the transverse field acting on the spin at site $i$. We assume 
$J_{ij}=J_{\rm s}$ if both spins belong to surface layers and $J_{ij}=J$ 
otherwise.

The statistical properties of the system are studied using an 
effective field theory whose the starting point is the generalized, but 
approximate, Callen \cite{32} relation derived by S\'{a} Barreto et al.~\cite{33} for 
the transverse Ising model. The longitudinal and transverse magnetizations of 
the spin at any site $i$ are approximately given by (for details see 
S\'{a} Barreto and Fittipaldi \cite{34})
\begin{eqnarray}
m_{i}^{z}&=&\left<\sigma_{i}^{z}\right>=\frac{1}{2}
\left<\frac{\sum_{j}J_{ij}\sigma_{j}^{z}}{[\Omega_{i}^{2}+
(\sum_{j}J_{ij}\sigma_{j}^{z})^{2}]^{\frac{1}{2}}}
\tanh\left(\frac{1}{2}\beta[\Omega_{i}^{2}+
(\sum_{j}J_{ij}\sigma_{j}^{z})^{2}]^{\frac{1}{2}}
\right)\right>\nonumber\\
	 &=&\left<f_{z}\left(\sum_{j}J_{ij}\sigma_{j}^{z},
	 \Omega_{i}\right)\right>;\label{eq2}\\
m_{j}^{x}&=&\left<\sigma_{i}^{x}\right>=\frac{1}{2}
\left<\frac{\Omega_{i}}{[\Omega_{i}^{2}+
(\sum_{j}J_{ij}\sigma_{j}^{z})^{2}]^{\frac{1}{2}}}
\tanh\left(\frac{1}{2}\beta[\Omega_{i}^{2}+
(\sum_{j}J_{ij}\sigma_{j}^{z})^{2}]^{\frac{1}{2}}
\right)\right>\nonumber\\
	 &=&\left<f_{x}\left(\sum_{j}J_{ij}
	 \sigma_{j}^{z},\Omega_{i}\right)\right>=
	    \left<f_{z}\left(\Omega_{i},\sum_{j}J_{ij}
	    \sigma_{j}^{z}\right)\right>;\label{eq3}
\end{eqnarray}
where $m_{i}^{z}$ and $m_{i}^{x}$ are respectively the longitudinal and
transverse magnetizations at site $i$, $\beta=1/k_{\rm B}T$ 
(we take $k_{\rm B}=1$ for simplicity), $\left<\ldots\right>$ indicates 
the usual canonical ensemble thermal average for a given configuration, 
and the sum runs over all nearest neighbors of site $i$. We assume that
the transverse field $\Omega$ depends only on the layer index, which 
we shall denote by $n$. Because of the translational symmetry parallel 
to the $(001)$ plane, also the magnetizations only depend on $n$.

To perform thermal averaging on the right-hand side of eqs.~(\ref{eq2},\ref{eq3}),
we follow the general approach described in ref.~\cite{30}. First of all, 
in the spirit of the effective field theory, multi-spin correlation functions 
are approximated by products of single spin averages. We then take advantage
of the integral representation of the Dirac's delta distribution, in order 
to write eqs.~(\ref{eq2},\ref{eq3}) in the form
\begin{equation}
m_{n}^{\alpha}=\int \d\omega\, f_{\alpha}(\omega,\Omega_{n})\,
\frac{1}{2\pi}\int
\d t\exp(\mathrm{i}\omega t)\prod_{j}\left<\exp(-\mathrm{i} tJ_{ij}
\sigma_{j}^{z})\right>,\label{eq4}
\end{equation}
where $\alpha=z,x$ and
\begin{eqnarray}
f_{z}(y,\Omega_{n})&=&\frac{1}{2}\frac{y}{[y^{2}+\Omega_{n}^{2}]^{\frac{1}{2}}}
\tanh\left(\frac{1}{2}\beta
[y^{2}+\Omega_{n}^{2}]^{\frac{1}{2}}\right),\label{eq5}\\
f_{x}(y,\Omega_{n})&=&\frac{1}{2}\frac{\Omega_{n}}{[y^{2}+
\Omega_{n}^{2}]^{\frac{1}{2}}}\tanh\left(\frac{1}{2}\beta
		      [y^{2}+\Omega_{n}^{2}]^{\frac{1}{2}}\right)\nonumber\\
		   &=&f_{z}(\Omega_{n},y).\label{eq6}
\end{eqnarray}

We now introduce the probability distribution of the spin variables 
(for details see Saber \cite{30} and Tucker et al.~\cite{31}):
\begin{equation}
P(\sigma_{n}^{z})=\frac{1}{2}\left[(1-2m_{n}^{z})\delta\left(\sigma_{n}^{z}
+\frac{1}{2}\right)+
	 (1+2m_{n}^{z})\delta\left(\sigma_{n}^{z}-\frac{1}{2}\right)\right].\label{eq7}
\end{equation}
Using this expression and eq.~(\ref{eq4}), we obtain the following set of equations 
for the longitudinal layer magnetizations:
\begin{eqnarray}
m_{1}^{z}&=&2^{-N-N_{0}}\sum_{\mu=0}^{N}\sum_{\mu_{1}=0}^{N_{0}}
	    C_{\mu}^{N}C_{\mu_{1}}^{N_{0}}
	    (1-2m_{1}^{z})^{\mu}(1+2m_{1}^{z})^{N-\mu}\nonumber\\
	 & &{}\times(1-2m_{2}^{z})^{\mu_{1}}(1+2m_{2}^{z})^{N_{0}-\mu_{1}}
	 \nonumber\\ & &{}\times 
	 f_{z}\left(\frac{J}{2}\left[R(N-2\mu)
	    +(N_{0}-2\mu_{1})\right],\Omega_{1}\right);\label{eq8}\\
m_{n}^{z}&=&2^{-N-2N_{0}}\sum_{\mu=0}^{N}\sum_{\mu_{1}=0}^{N_{0}}
	    \sum_{\mu_{2}=0}^{N_{0}}C_{\mu}^{N}C_{\mu_{1}}^{N_{0}}
	    C_{\mu_{2}}^{N_{0}}(1-2m_{n}^{z})^{\mu}(1+2m_{n}^{z})^{N-\mu}\nonumber\\
	 & &{}\times(1-2m_{n-1}^{z})^{\mu_{1}}(1+2m_{n-1}^{z})^{N_{0}-\mu_{1}}
	    (1-2m_{n+1}^{z})^{\mu_{2}}(1+2m_{n+1}^{z})^{N_{0}-\mu_{2}}\nonumber\\
	 & &{}\times f_{z}\left(\frac{J}{2}\left[(N+2N_{0})-2(\mu+
	    \mu_{1}+\mu_{2})\right],\Omega_{n}\right),\nonumber\\
	   && \qquad\hbox{\rm for\ }n=2,3,\ldots,L-1;\label{eq9}\\
m_{L}^{z}&=&2^{-N-N_{0}}\sum_{\mu=0}^{N}\sum_{\mu_{1}=0}^{N_{0}}
	    C_{\mu}^{N}C_{\mu_{1}}^{N_{0}}
	    (1-2m_{L}^{z})^{\mu}(1+2m_{L}^{z})^{N-\mu}
	    \nonumber\\	 & &{}\times
	    (1-2m_{L-1}^{z})^{\mu_{1}}(1+2m_{L-1}^{z})^{N_{0}-\mu_{1}}
	 \nonumber\\	 & &
	{}\times f_{z}\left(\frac{J}{2}\left[R(N-2\mu)
	    +(N_{0}-2\mu_{1})\right],\Omega_{L}\right).\label{eq10}
\end{eqnarray}
In these equations we have introduced the notation $R=J_{\rm s}/J$, 
$N$ and $N_{0}$ are the number of nearest neighbors in the plane and 
between adjacent planes respectively, and $C_{k}^{l}$ are
the binomial coefficients, $C_{k}^{l}={l!}/({k!(l-k)!})$.
For the case of a simple cubic lattice, one has $N=4$ and $N_{0}=1$.

The equations for the transverse magnetization for each layer are 
obtained by substituting the function $f_{x}$ instead of $f_z$ in the 
expression of the longitudinal magnetization. Since however 
$f_x(y,\Omega)=f_z(\Omega,y)$, this yields
\begin{equation}
m_{n}^{x}=m_{n}^{z}[f_{z}(y,\Omega_{n}) \longrightarrow f_{x}(y,\Omega_{n})]
	 =m_{n}^{z}[f_{z}(\Omega_{n},y)].\label{eq11}
\end{equation}
We have thus obtained a set of self consistent equations (\ref{eq8}--\ref{eq11}) for the layer
longitudinal and transverse magnetizations $m_{n}^{z}$, $m_n^x$, that can be 
directly solved by numerical iteration. However, since we are interested in
the calculation of the longitudinal order near the critical temperature,
the usual argument that the layer longitudinal magnetizations $m_{n}^{z}$
should tend to zero as the temperature approaches its critical value, allows us 
to consider only terms linear in $m_{n}^{z}$, because higher order terms 
tend to zero faster than $m_{n}^{z}$. 
Consequently, all terms of order higher than linear 
in eqs.~(\ref{eq8}--\ref{eq11}) can be 
neglected. This leads to the following system of equations:
\begin{equation}
m_{n}^{z}=A_{n,n-1}m_{n-1}^{z}+A_{nn}m_{n}^{z}+A_{n,n+1}m_{n+1}^{z},\label{eq12}
\end{equation}
which can be written as
\begin{equation}
{\sf A}\,m_{n}^{z}=m_{n}^{z},\label{eq13}
\end{equation}
where the matrix ${\sf A}$ is symmetric and tridiagonal with elements
\begin{equation}
A_{ij}=A_{ii}\delta_{ij}+A_{ij}(\delta_{i,j-1}+\delta_{i,j+1}).
\end{equation}
For simplicity, we now assume that the transverse field acting on the system 
is uniform and equal to $\Omega$. In this case, the only nonzero elements of
the matrix ${\sf A}$ are given by
\begin{eqnarray}
A_{11} &=&A_{LL}=\frac{1}{2}\left\{f_{z}\left(\frac{J}{2}(4R+1),\Omega\right)+
	    f_{z}\left(\frac{J}{2}(4R-1),\Omega\right)\right.\nonumber\\
	 & &\left.{}+2f_{z}\left(\frac{J}{2}(2R+1),\Omega\right)+
	    2f_{z}\left(\frac{J}{2}(2R-1),\Omega\right)\right\};\\
A_{12}  &=&A_{L,L-1}=\frac{1}{8}\left\{f_{z}\left(\frac{J}{2}(4R+1),
	\Omega\right)-
	    f_{z}\left(\frac{J}{2}(4R-1),\Omega\right)\right.\nonumber\\
	 & &\left.{}+4f_{z}\left(\frac{J}{2}(2R+1),\Omega\right)-
	    4f_{z}\left(\frac{J}{2}(2R-1),\Omega\right)+6
	    f_{z}\left(\frac{J}{2},\Omega\right)\right\};\\
A_{nn}  &=&4A_{n,n-1}=4A_{n,n+1}=
\frac{1}{4}\left\{f_{z}(3J,\Omega)+4f_{z}(2J,\Omega)+
	    5f_{z}(J,\Omega)\right\};\nonumber\\
	    &&\qquad\hbox{for\ \ }n=2,3,\ldots,L-1.
\end{eqnarray}
The system of eqs.~(\ref{eq12}) is of the form
\begin{equation}
{\sf M}\,m_{n}^{z}=0,\label{eq18}
\end{equation}
where the elements of the matrix ${\sf M}$ are given by
\begin{equation}
M_{ij}=(1-A_{ii})\delta_{ij}-A_{ij}(\delta_{i,j-1}+\delta_{i,j+1}).
\end{equation}
All the information about the critical temperature of the system is contained
in eq.~(\ref{eq18}). So far we have not assigned explicite values to the coupling
constants and the transverse field: the terms in matrix (\ref{eq18}) are general ones.

In a general case, for arbitrary coupling constants, transverse 
field and film thickness, the evaluation of the critical temperature relies on 
numerical solution of the system of linear equations (\ref{eq18}). These equations 
can be satisfied by nonzero magnetization vectors $m_n^{z}$ only if
\begin{equation}
\mathop{\mathrm{Det}}{\sf M}=0,\label{eq20}
\end{equation}
where
\begin{equation}
\mathop{\mathrm{Det}}{\sf M}=c\left|
\begin{array}{ccccc}
 a & -1 &  &  &   \\
-1 & b & -1 &  &    \\ 
 \ddots& \ddots & \ddots & \ddots &\\ 
& -1 & b & -1 &   \\ 
&\ddots & \ddots & \ddots & \ddots  \\
  &  & -1 & b & -1 \\ 
  &  &  & -1 & a
\end{array}
\right|_{L}.\label{eq21}
\end{equation}
The parameters $a$, $b$ and $c$ that appear in equation (\ref{eq21}) take 
into account the boundary conditions and represent the different propensities 
to order of the surfaces and of the bulk. They are given by
\begin{eqnarray}
a &=&\frac{1-A_{11}}{A_{12}};\label{eq22}\\
b &=&\frac{1-A_{nn}}{A_{n,n-1}}=\frac{1-A_{nn}}{A_{n,n+1}}=
     \frac{1-A_{nn}}{A_{nn}/4};\qquad\hbox{\rm for\ \ }n=2,3,\ldots,L-1;\label{eq23}\\
c &=&\left(\frac{1}{A_{12}}\right)^{2}\left(\frac{1}{A_{nn}/4}\right)^{2(L-2)}.\label{eq24}
\end{eqnarray}
In general, equation (\ref{eq20}) can be satisfied for $L$ different values of the 
critical temperature $T_{\rm c}/J$ from which we choose the one corresponding 
to the highest possible transition temperature (cfr.\ the discussion in 
refs.~\cite{34,35}). This value of $T_{\rm c}/J$ corresponds to a solution
where $m_{1}^{z}$, $m_{2}^{z}$,\dots, $m_{L}^{z}$ are all positive, which is 
compatible with a ferromagnetic longitudinal ordering. The other solutions 
correspond in principle to other types of ordering that usually do not occur
here~\cite{34}.

The reduction and rearrangement of the determinant of eq.~(\ref{eq21}) 
leads to the result \cite{37,38,39}
\begin{equation}
\mathop{\mathrm{Det}}{\sf M}=c[(ab-1)^{2}D_{L-4}b-2a(ab-1)D_{L-5}b+a^{2}D_{L-6}b],
\label{eq25}
\end{equation}
where $D_{L}(x)$ is the determinant
\begin{equation}
D_{L}(x)=\left|
\begin{array}{ccccc}
 x & -1 &  &  &    \\
-1 & x & -1 &  &    \\
\ddots & \ddots & \ddots & \ddots &  \\ 
& -1 & x & -1 &   \\
 & \ddots & \ddots & \ddots & \ddots \\
  &  & -1 & x & -1 \\
  &  &  & -1 & x
\end{array}
\right|_{L},
\end{equation}
whose value is given by
\begin{eqnarray}
D_{L}(x)&=&(x^{2}-4)^{-\frac{1}{2}}\,2^{-(L+1)}\nonumber\\
	 &&{}\times\left\{\left[x+\sqrt{x^{2}-4}\right]^{L+1}
	 -\left[x-\sqrt{x^{2}-4}\right]^{L+1}\right\},
	 \quad\hbox{for }x^2>4;\\
&=&\sin[(L+1)k]/\sin k,\ \hbox{\rm with\ }k=\cos^{-1}(x/2),
\qquad\hbox{for }x^2\le 4.
\end{eqnarray}
From now on, we take $J$ as the unit of energy in our numerical 
calculations, and we measure length in units of the lattice constant.

\section{Phase diagrams}
From eqs.~(\ref{eq20}) and (\ref{eq25}), we can obtain the phase diagrams of the film. The
results show that there can be two phases, a film ferromagnetic phase (F),
in which the longitudinal magnetization
(${\overline{m}_{z}}=\frac{1}{L}\sum_{n=1}^{L}m_{n}^{z}$) is different
from zero, and a film paramagnetic phase (P), in which
${\overline{m}_{z}}=0$. In addition, if the number of layers in the film, $L$, 
is very large ($L \rightarrow \infty$), the film should
practically behave as a semi-infinite Ising system. It is well known that, if the
ratio $R=J_{\rm s}/J$ of surface to bulk exchange interactions in a semi-infinite Ising system,
is greater than a critical value 
$R_{\rm c}=(J_{\rm s}/J)_{\rm crit}$, there appear two different transitions: 
the {\em surface\/} transition and the
{\em bulk\/} transition. The critical temperatures related to them are respectively called
the {\em surface\/} critical temperature $T_{\rm c}^{\rm S}$ and the {\em bulk\/} critical 
temperature $T_{\rm c}^{\rm B}$. 
To obtain the bulk and surface critical temperatures of the semi-infinite
Ising system, we follow the approach due to Binder and Hohenberg \cite{40}.
Using eqs.~(\ref{eq22}--\ref{eq24}), the system of linear equations (\ref{eq13}) yields
\begin{eqnarray}
am_{1}^{z}-m_{2}^{z} &=& 0;\label{eq28}\\
-m_{1}^{z}+bm_{2}^{z}-m_{3}^{z}&=&0;\label{eq29}\\
-m_{n-1}^{z}+bm_{n}^{z}-m_{n+1}^{z}&=&0,\qquad \hbox{for }n\geq 3.\label{eq30}
\end{eqnarray}
According to Binder and Hohenberg \cite{40}, let us assume that 
$m_{\rm n+1}^{z}=\gamma \ m_{\rm n}^{z}$ for $n\geq 3$, e.g., the layer 
magnetization $m_{\rm n}^{z}$ of each layer, with $n$ larger than $2$, 
decreases exponentially in the bulk. The equations (\ref{eq28}) 
and (\ref{eq29}) then yield the following secular equation:
\begin{equation}
M_s \left( 
\begin{tabular}{c}
$m_{1}^{z}$ \\ 
$m_{2}^{z}$
\end{tabular}
\right) =\left( 
\begin{tabular}{cc}
$a$  & $-1$ \\ 
$-1$ & $b-\gamma $
\end{tabular}
\right) \left( 
\begin{tabular}{c}
$m_{1}^{z}$ \\ 
$m_{2}^{z}$
\end{tabular}
\right) =0,\label{eq31}
\end{equation}
where, from eq.~(\ref{eq30}), the parameter $\gamma $ is given by
\begin{equation}
\gamma =\frac{1}{2}\left( b -\sqrt{b^2-4}\right).\label{eq33}
\end{equation}
Thus, the surface critical temperature $T_c^S/J$ can be derived from the
condition $\mathop{\mathrm{Det}}M_s=0$, namely 
\begin{equation}
a\left( b-\gamma \right) - 1=0.\label{eq34}
\end{equation}
We can now study numerically the physical properties of the surface
and the bulk of the semi-infinite Ising system. Here it is worth
noting that in our treatment the bulk transition temperature can be
determined by letting $m_{\rm n}^{z}=m_{\rm n-1}^{z}=m_{\rm n+1}^{z}=m^z$ in 
eq.~(\ref{eq30}), i.e.,
\begin{equation}
b-2=0.
\end{equation}
This yields 
\begin{equation}
f_z(3J,\Omega) +4f_z( 2J,\Omega) +5f_z(J,\Omega) =8/3  .\label{eq35}
\end{equation}
At $T_c^B/J=0$, eq.~(\ref{eq35}) yields the bulk critical 
transverse field value: $\Omega_{\rm c}^{B}/J=2.3529$.
On the other hand, for the special case of the Ising model in the
absence of the transverse field ($\Omega=0$), eq.~(\ref{eq35}) reduces to 
\begin{equation}
\tanh (3\beta J)+4\tanh (2\beta J)+5\tanh (\beta J)=16/3,
\end{equation}
which is the Zernike \cite{41} equation for the simple cubic lattice. The
transition temperature is then determined as $T_c^B/J=1.2683$.

\begin{figure}
\begin{center}
\epsfig{file=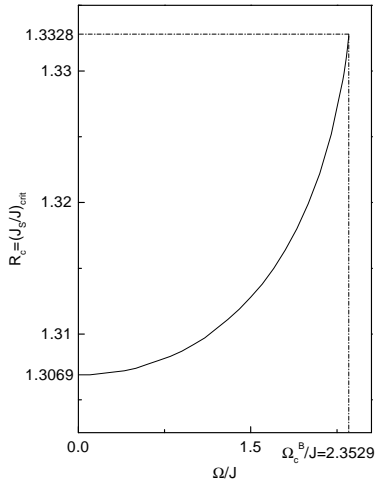,width=6cm}
\end{center}
\caption{The variation of $R_{\rm c}=(J_{\rm s}/J)_{\rm crit}$ 
as a function of the transverse field $\Omega/J$.}\label{fig1}
\end{figure}

A useful expression for determining the critical value $R_c=(J_s/J)_{\rm
crit}$ is therefore given by the simultaneous solution of the equations
(\ref{eq33}) and (\ref{eq34}). The variation of $R_{\rm c}$ as a
function of the transverse field is shown in Fig.~\ref{fig1}. It shows
that $R_{\rm c}$ increases with the increase of the strength of the
transverse field from its minimal value $R_{\rm c}^{\rm min}=1.3069$ for
$\Omega/J=0$ and reaches its maximal value $R_{\rm c}^{\rm max}=1.3328$
at the bulk critical transverse field $\Omega_{\rm c}^{\rm B}/J=2.3529$.

\begin{figure}
\begin{center}
\epsfig{file=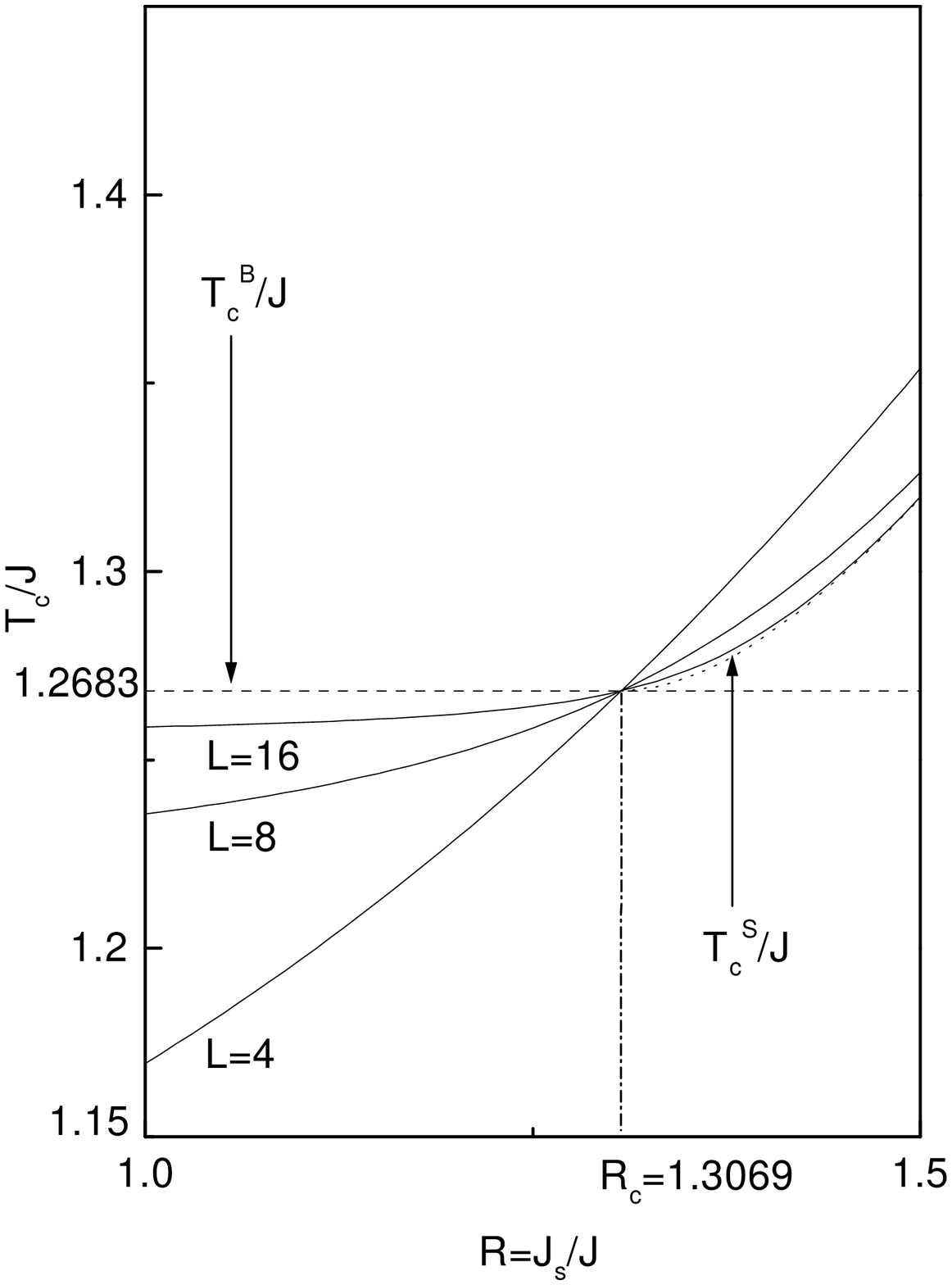,width=6cm}
\epsfig{file=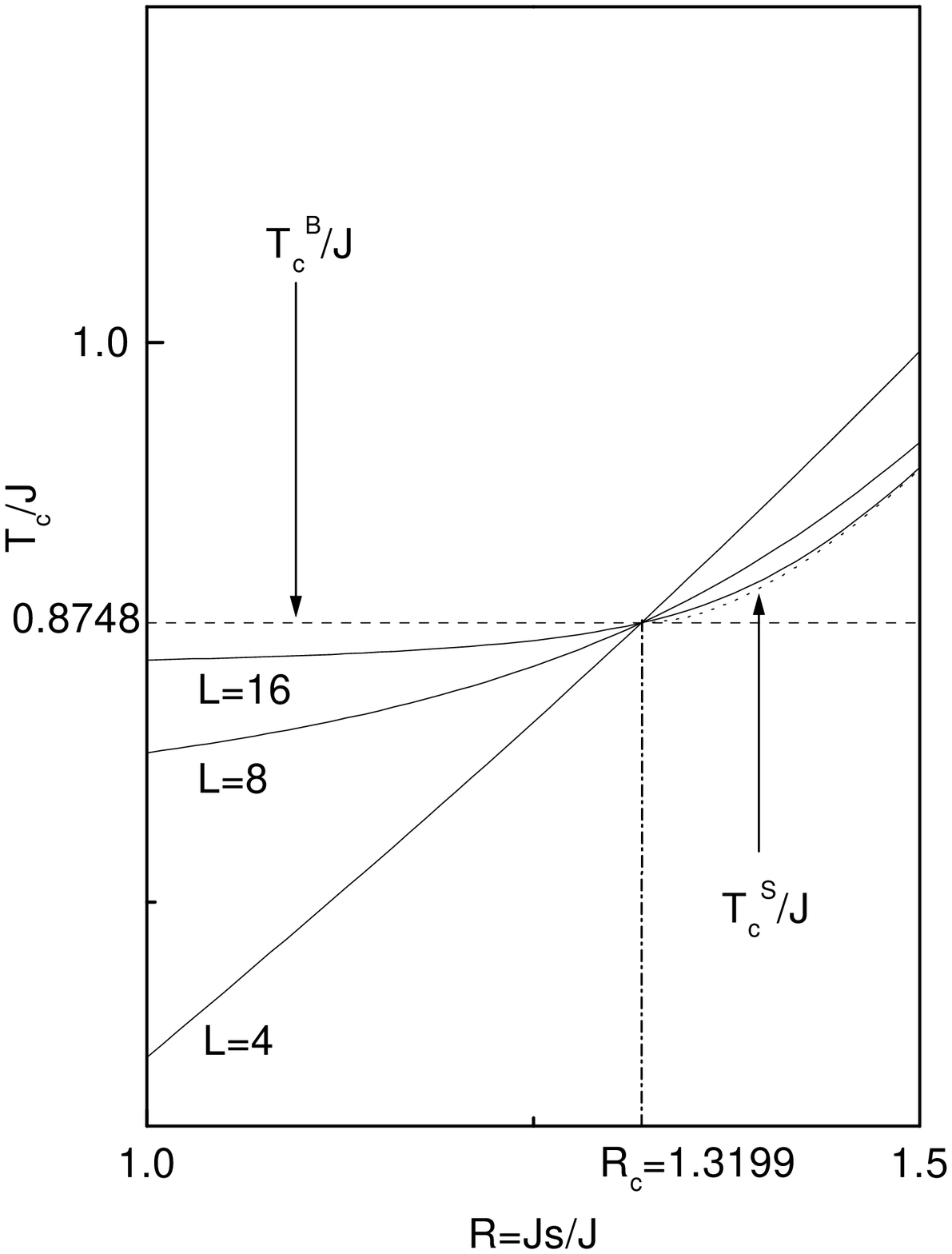,width=6cm}
\end{center}
\caption{The phase diagram in the $(T_{\rm c}/J,R=J_{\rm s}/J)$ 
plane.
(a) $\Omega/J=0 $;
(b) $\Omega/J=2 $.
The dashed line is the bulk critical temperature $T_{\rm c}^{\rm B}/J$ and the
dotted line is the surface critical temperature $T_{\rm c}^{\rm S}/J$ of the
corresponding semi-infinite Ising system.}\label{fig2}
\end{figure}

We now calculate the $(T_{\rm c}/J, R=J_{\rm s}/J)$ phase diagrams for
different values of the transverse field and of the number of layers:
typical results are shown in Fig.~\ref{fig2}. Our phase diagrams are
qualitatively different from the corresponding diagrams for the
semi-infinite ferromagnet. The main difference is that we get in the
film only one well defined critical temperature $T_{\rm c}/J$, instead
of the two critical temperatures $T_{\rm c}^{\rm B}/J$ and $T_{\rm
c}^{\rm S}/J$. This temperature depends on the film thickness. According
to these results we must give a new definition of  $R_{\rm c}$. In
semi-infinite systems $R_{\rm c}$ was defined as the value of the
parameter $R$ above which the two critical temperatures $T_{\rm c}^{\rm
B}/J$ and $T_{\rm c}^{\rm S}/J$ exist. However, according to
Fig.~\ref{fig2} the parameter $R_{\rm c}$ can now be defined as that
particular value of $R$ at which the critical temperature does not
depend on film thickness (the cross-over point in Fig.~\ref{fig2}). As
it has been stated the numerical values of $R_{\rm c}$ and related
$T_{\rm c}/J$ parameters are exactly the same as those found for the
semi-infinite system. Furthermore, according to the definition of
$R_{\rm c}$, it can be expected that the cross-over point in
Fig.~\ref{fig2} should also define the critical temperature of the
three-dimensional infinite bulk system, where the surfaces and the $R$
parameter are of no importance. Fig.~\ref{fig2}, where the bulk and the
surface critical temperatures of the corresponding semi-infinite system
are represented respectively by the dashed and dotted lines also shows
that this is really the case.

From Fig.~\ref{fig2}a, which corresponds to the case when there is no
transverse field acting on the system, we find that the value of $R_{\rm
c}$ corresponding to the cross-over point is equal to $1.3069$, which is
equal to the value reported by Wiatrowski et al.~[42] and by Sarmento
and Tucker [43]. For $R<R_{\rm c}$, the critical temperature $T_{\rm
c}/J$ of the film is smaller than the bulk critical temperature $T_{\rm
c}^{\rm B}/J$, and $T_{\rm c}/J$ increases with the increase of $L$,
approaching the bulk critical temperature $T_{\rm c}^{\rm B}/J=1.2683$
asymptotically as the number of layers becomes large. When $R=R_{\rm
c}$, the critical temperature of the film $T_{\rm c}/J$ is independent
of $L$, and equal to $T_{\rm c}^{\rm B}/J$. On the other hand, for
$R>R_{\rm c}$ the critical temperature of the film $T_{\rm c}/J$ is
larger than both the bulk $T_{\rm c}^{\rm B}/J$ and the surface
$T_{\rm c}^{\rm S}/J$ critical temperatures of the corresponding
semi-infinite system. The larger $L$, the lower $T_{\rm c}/J$, and, when
the number of layers $L$ becomes large, $T_{\rm c}^{\rm B}$ approaches
asymptotically $T_{\rm c}^{\rm S}/J$.

We now study the influence of the surface exchange interactions in the
presence of a transverse field. Fig.~\ref{fig2}b shows that a transverse
field acting on the system increases the critical value of $R_{\rm c}$.
For example, with a transverse field $\Omega/J=2$, the critical value of
$R_{\rm c}$ is shifted from 1.3069 to 1.3199. At the same time of
course for a given ratio of the exchange interactions $R=J_{\rm s}/J$,
the critical temperatures of the film and of the corresponding
semi-infinite system are reduced.

\begin{figure}
\begin{center}
\epsfig{file=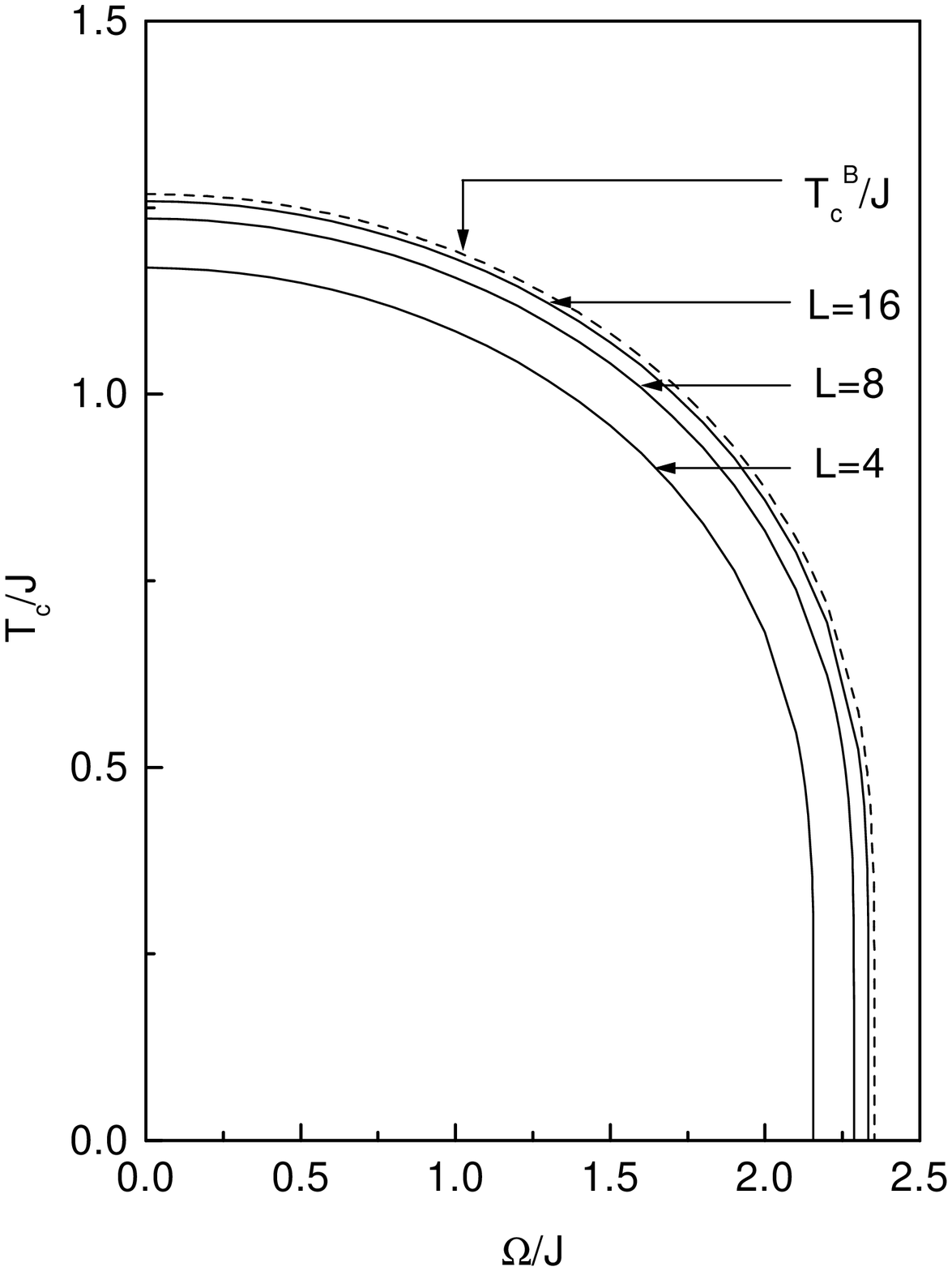,width=6cm}
\epsfig{file=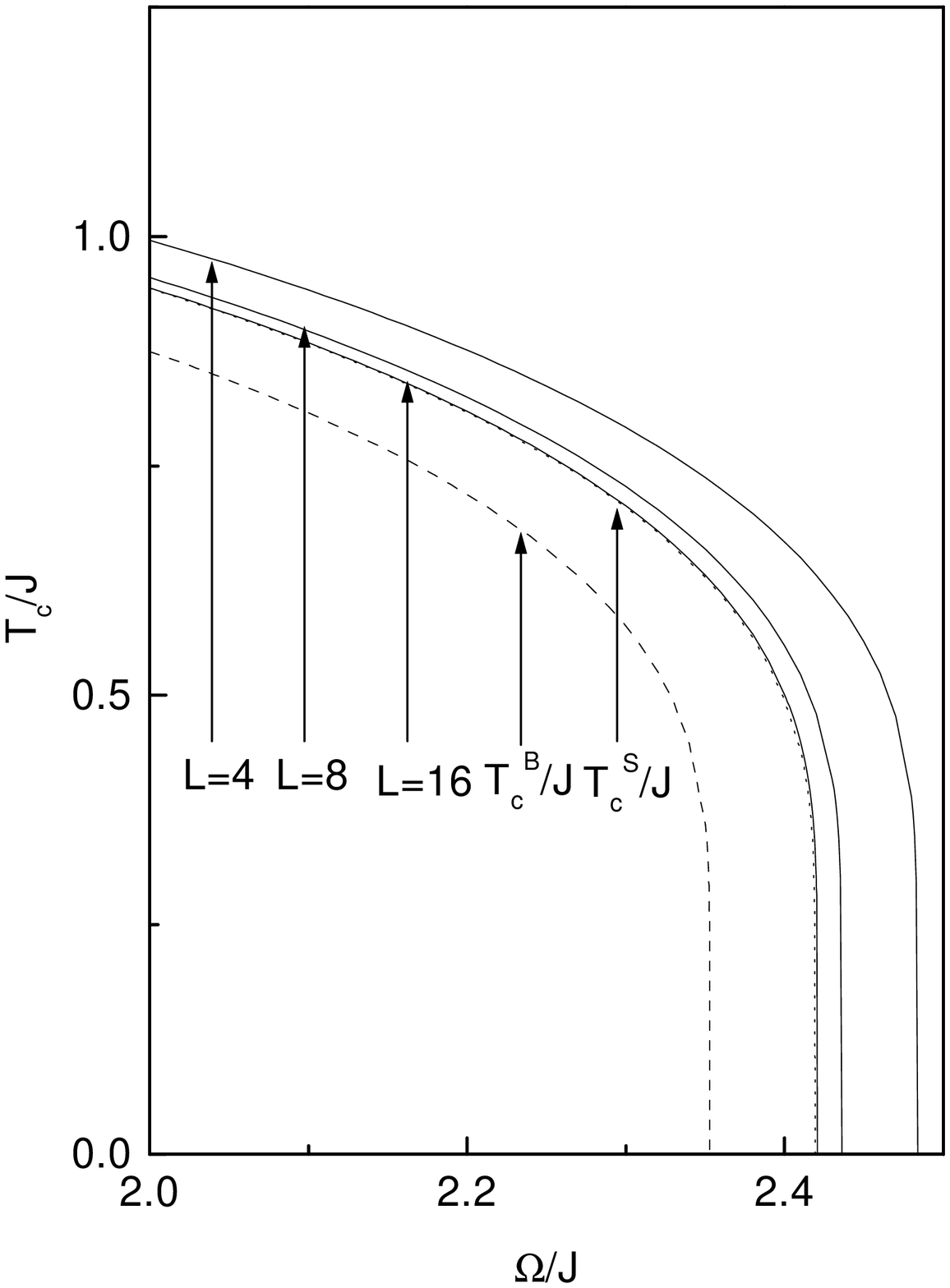,width=6cm}
\end{center}
\caption{Phase diagram in the $(T_{\rm c}/J,\Omega/J)$ plane:
(a) $R=1$; (b) $R=1.5$.
The dashed line is the bulk critical temperature $T_{\rm c}^{\rm B}/J$ and the
dotted line is the surface critical temperature $T_{\rm c}^{\rm S}/J$ of the
corresponding semi-infinite Ising system.}\label{fig3}
\end{figure}

In Figs.~\ref{fig3}, we show the phase diagrams of the film in the
($T_{\rm c}/J,\Omega/J$)-plane with different values of the thickness
$L$, and the critical temperatures of the semi-infinite Ising system for
different values of the parameter $R$. The presence of a transverse
field, of course, reduces the critical temperatures of the film and of
the semi-infinite system. We find that the $(T_{\rm c}/J,\Omega/J)$
curve for a given value of $R$ intercepts the $(\Omega/J)$-axis at a
critical value $\Omega_{\rm c}/J$ of the transverse field. When
$\Omega/J>\Omega_{\rm c}/J$, there cannot be a ferromagnetic phase at
any temperature. Fig.~\ref{fig3}a, which corresponds to $R=1$ (smaller
than $R_{\rm c}^{\rm min}$), shows that, for any finite value of the
film thickness $L$, the film critical temperature $T_{\rm c}/J$ is
smaller than the bulk critical temperature $T_{\rm c}^{B}/J$ and
increases as $L$ increases, approaching $T_{\rm c}^{\rm B}/J$ for large
values of $L$. Fig.~\ref{fig3}b corresponds to $R=1.5>R_{\rm c}^{\rm
min}$, and shows that the film critical temperature $T_{\rm c}/J$ is
larger than both the bulk critical temperature $T_{\rm c}^{\rm B}/J$ and
the surface critical temperature $T_{\rm c}^{\rm S}/J$ of the
corresponding semi-infinite system. $T_{\rm c}/J$ decreases with the
increase of $L$, approaching $T_{\rm c}^{\rm S}/J$ for large values of
$L$.

\begin{figure}
\begin{center}
\epsfig{file=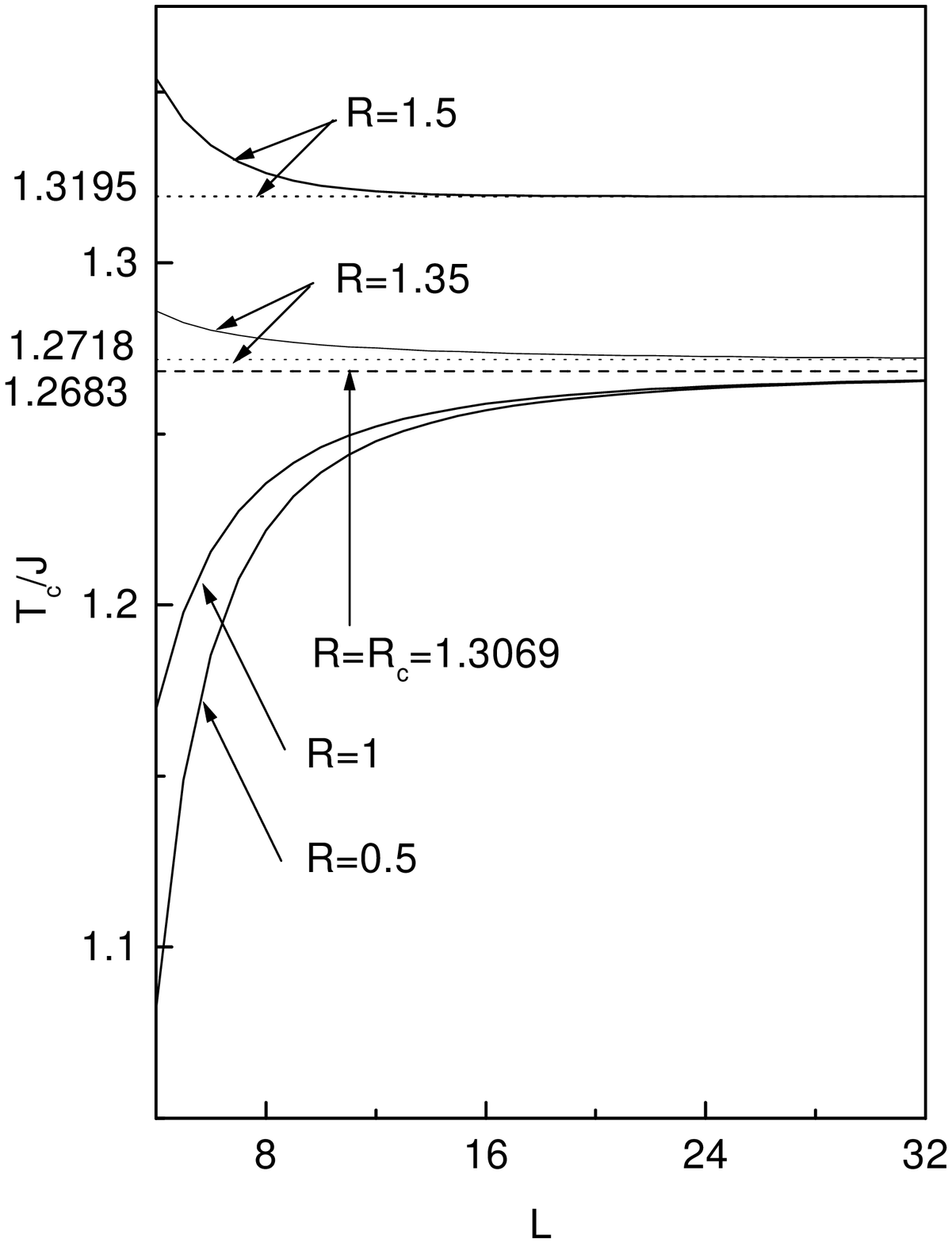,width=6cm}
\epsfig{file=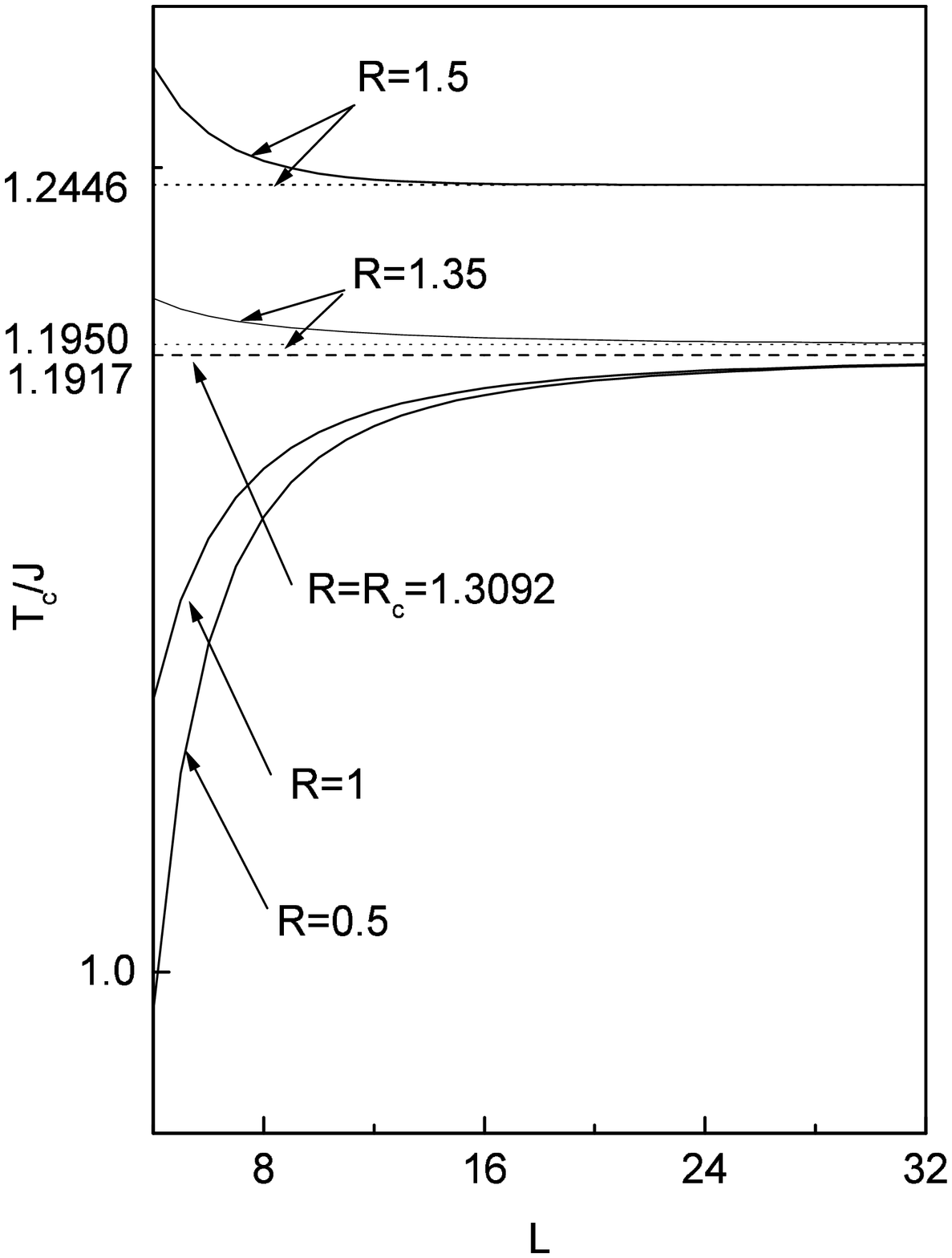,width=6cm}
\end{center}
\caption{Thickness dependence of the critical temperature of the
film for (a) $\Omega/J=0$, and (b) $\Omega/J=1$.
The dashed line is the bulk critical temperature $T_{\rm c}^{\rm B}/J$ and the
dotted line is the surface critical temperature $T_{\rm c}^{\rm S}/J$ of the
corresponding semi-infinite Ising system.}\label{fig4}
\end{figure}

We show in Fig.~\ref{fig4} the thickness dependence of the critical
temperature of the film for different values of $R$ and $\Omega/J$.
Fig.~\ref{fig4}a corresponds to the case when the transverse field
vanishes. It shows that for any value of $R$ below the critical value
$R_{\rm c}=R_{\rm c}^{\rm min}=1.3069$, the critical temperature of the
film increases with $L$ and approaches the bulk critical temperature
$T_{\rm c}^{\rm B}/J~=~1.2683$ asymptotically as the number of layers
becomes large. On the other hand, for $R>R_{\rm c}$, the critical
temperature of the film becomes smaller as the number of layers
increases, and approaches asymptotically, for large values of $L$, the
surface critical temperature $T_{\rm c}^{\rm S}/J$ which depends on $R$.
In Fig.~\ref{fig4}b, we show the thickness dependence of the critical
temperature of the film when there is a transverse field acting on the
system. We consider the case when $\Omega/J=1$ for several values of
$R$. We see that Fig.~\ref{fig4}b exhibits the same qualitative behavior
as Fig.~\ref{fig4}a, except that the transverse field reduces the
critical temperatures.

\begin{figure}
\begin{center}
\epsfig{file=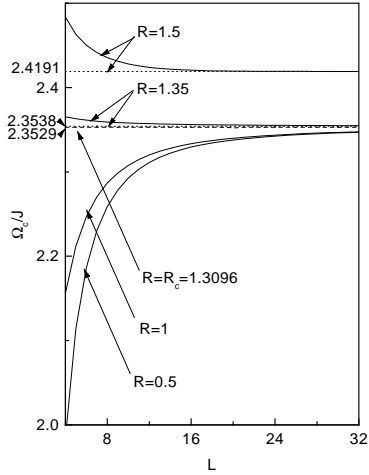,width=6cm}
\end{center}
\caption{The variation of the critical transverse field of the film
$\Omega_{\rm c}/J$ at which the critical temperature of the film $T_{\rm c}/J$
becomes zero as function of the number of layers $L$.
The dashed line and the dotted line correspond respectively to the bulk and
surface critical fields of the semi-infinite Ising system.}\label{fig5}
\end{figure}

Finally Fig.~\ref{fig5} shows the variation of the critical transverse field
$\Omega_{\rm c}/J$ as a function of the thickness of the film $L$ for
several values of $R$. The dashed and dotted lines correspond respectively
to the bulk and surface critical transverse fields of the corresponding
semi-infinite Ising system. For $R\le R_{\rm c}$ the critical transverse
field of the film, $\Omega_{\rm c}/J$,  is smaller than the bulk critical
transverse field $\Omega_{\rm c}^{\rm B}/J$, and increases with the
increase of $L$,  approaching $\Omega_{\rm c}^{\rm B}/J$ for large values
of $L$. For $R>R_{\rm c}$, $\Omega_{\rm c}/J$ is larger both than 
$\Omega_{\rm c}^{\rm B}/J$ and $\Omega_{\rm c}^{\rm S}/J$, and decreases
as $L$ increases, approaching $\Omega_{\rm c}^{\rm S}/J$ for large 
values of $L$.

\section{Conclusion} 
We have studied, within the effective field theory,
the phase diagram of the transverse
spin-$\scriptstyle{\frac{1}{2}}$ Ising film, where the exchange
interactions between spins on the surfaces are different from those in the bulk. 
We have investigated
the effects of the surface to bulk exchange interaction ratio, of the
strength of the transverse field and of the film thickness on the phase
diagram. The results show that, in the film, there is only one critical
temperature $T_{\rm c}/J$ which depends on $L$, $R$ and $\Omega/J$. We
have identified a critical value $R_{\rm c}$ of the parameter $R$, such
that when $R=R_{\rm c}$, $T_{\rm c}/J$ is independent of $L$; for $R\le
R_{\rm c}$, $T_{\rm c}/J$ is smaller than $T_{\rm c}^{\rm B}/J$ and for
$R>R_{\rm c}$, $T_{\rm c}/J$ is greater both than $T_{\rm c}^{\rm B}/J$
and $T_{\rm c}^{\rm S}/J$. When $L$ becomes very large, then for $R\le
R_{\rm c}$ ($R>R_{\rm c}$), $T_{\rm c}/J$ approaches $T_{\rm c}^{\rm
B}/J$ ($T_{\rm c}^{\rm S}/J$) of the corresponding semi-infinite Ising
system.

\begin{ack} 
This work began during a visit of MS to the
Dipartimento di Scienze Fisiche, Universit\`{a} ``Federico II'', in the
framework of an exchange programme between the CNR (Italy) and the
CNCPRST (Morocco) and was completed during a visit of both authors to the
AS-ICTP, Trieste, Italy. The authors would like to thank all these
organizations. 
\end{ack}

\end{document}